\documentclass[prd,preprint,a4paper,superscriptaddress,showpacs]{revtex4-1}

\usepackage{graphicx, subfigure}
\usepackage{epstopdf}

\usepackage[colorlinks]{hyperref}
\newcommand\aj{Astronomical Journal}
\newcommand\araa{Annual Review of Astron and Astrophys}
\newcommand\aap{A\&A}
\newcommand\jcap{JCAP}
\newcommand\mnras{MNRAS}
\newcommand\apjs{Astrophysical Journal Supplement}

\begin{document}

\title{Spherical `Top-Hat' Collapse in general Chaplygin gas dominated
universes}

\author{R. A. A. Fernandes}
\email[Electronic address: ]{Rui.Fernandes@astro.up.pt}
\affiliation{Centro de Astrof\'{\i}sica, Universidade do Porto, Rua das
Estrelas, 4150-762 Porto, Portugal}
\affiliation{Dipartimento di Scienze Matematiche, Fisiche e Chimiche,
Universit\`a dell'Insubria, Via Valleggio 11, 22100 Como, Italy}

\author{J. P. M. de Carvalho}
\affiliation{Departamento de Matemática da Faculdade Ci\^encias da Universidade
do Porto,Rua do Campo Alegre,
4169-007 Porto, Portugal
}
\affiliation{Centro de Astrof\'{\i}sica, Universidade do Porto, Rua das
Estrelas, 4150-762 Porto, Portugal}

\author{A. Yu. Kamenshchik}
\affiliation{Physics Department and INFN, University of
Bologna, via Irnerio 46, 40126 Bologna, Italy}
\affiliation{L.D. Landau Institute for Theoretical Physics of the Russian
Academy of Sciences, Kosygina str. 2, 119334 Moscow, Russia}

\author{U. Moschella}
\affiliation{Dipartimento di Scienze Matematiche, Fisiche e Chimiche,
Universit\`a dell'Insubria, Via Valleggio 11, 22100 Como, Italy}
\affiliation{INFN, Sez di Milano}

\author{A. da Silva}
\affiliation{Centro de Astrof\'{\i}sica, Universidade do Porto, Rua das
Estrelas, 4150-762 Porto, Portugal}

\date{\today}
\begin{abstract}
We expand previous works on the spherical `top-hat' collapse (SC-TH)
framework in generalized Chaplygin gas (gCg) dominated universes. Here we allow
the collapse in all energetic components within the system. We analyze the
non-linear stages
of collapse for various choices of parameter $\alpha$ of the gCg model
introducing an exact formulation for the so-called effective sound speed,
$c_{eff}^2$. We show that, within the SC-TH framework, the growth of the
structure becomes faster
with increasing values of $\alpha$. 
%This partially contradicts previous results
%obtained
%using an approximation in the context of linear perturbation theory.
\end{abstract}


%\date{\today}

\maketitle

\section{Introduction} \label{sec:Introduction}

More than a decade has passed since Type Ia Supernovae observations provided a
first indication of the present accelerated expansion of the Universe
\cite{perlmutter:1999,riess:1998}. The culprit, Dark Energy (DE) - in the form
of a cosmological constant - soon became the simplest way to describe this
dynamical behavior. Since then, the amount of observational evidence supporting
DE has grown in a way
that it has become a major player in the concordance model of cosmology (for a
review on DE see e.g.\ \cite{frieman:2008,caldwell:2009}). Although DE accounts
for
$\sim$70\% of the
present energy density of the Universe, its physical nature remains
undisclosed. Another key player of the concordance model, whose nature is also
unknown to particle physics, is dark matter (DM). Together with DE, they account
for most (95\%) of the present matter-energy density (the so called {\it dark
sector}) of the Universe. While DE may be modeled as a fluid with negative
pressure acting against gravitational collapse, DM (in its cold version) is a
dust like fluid with no
pressure, therefore enhancing the collapse of matter perturbations (for a review
on DM see e.g.\ \cite{feng:2010,garrett:2011}).

Although the cosmological constant is the simplest DE model that fits present
astronomical data, understanding its nature poses most pressing challenges to
particle physics and cosmology. Alternative DE models have been
proposed to alleviate some of the problems of the cosmological constant
concordance model. This include dynamical DE, minimally coupled or with
interactions (see e.g.\ \cite{gumjudpai:2005,copeland:2006}), and Unified Dark
Models
(UDM) (see e.g.\ \cite{avelino:2008, bertacca:2010}). In the later case, DE and
DM are
described by the same physical entity.
A particular UDM, first introduced in \cite{Kamenshichik:2001} and
subsequently developed by \cite{bilic:2002,bento:2002,gorini:2003}, is the
so-called {\em generalized Chaplygin gas} (gCg), which is based on the following
exotic equation of state (EoS):
\begin{equation}
p=-\frac{C}{\rho^{\,\alpha}},
\label{eq:EoS_gCg}
\end{equation}
where $p$ is the pressure, $\rho$ is the density, and $C$ and $\alpha$ are
constants (in general, both assumed to be positive). The $\alpha = 1$ case,
corresponds to the standard Chaplygin gas, named after the Russian
physicist Sergey A.\ Chaplygin who studied it in a
hydrodynamical context \cite{chaplygin:1901}.

Using Eq.~(\ref{eq:EoS_gCg}) together with the relativistic energy-momentum
conservation equation, one can show that gCg's background density evolution is
(see e.g.\ \cite{avelino:2003}):
\begin{equation}
 \rho = \rho_0 \left(\bar C + (1-\bar C)a^{-3(\alpha + 1)} \right)^{\frac{1}{1+
\alpha}}.
\label{eq:density_gCg}
\end{equation}
where $\bar C = C/\rho_0^{1+\alpha}$, $\rho_0$ is the density at the present
epoch and $a$ is the cosmic scale factor,
which is related to the cosmological redshift by $1+z=1/a$ (assuming the scale
factor normalized for the present epoch, i.e.\ $a_0 = 1$).

It can be easily shown that the EoS parameter, $w = p/\rho$, is given by:
\begin{equation}
w = -\bar C\left(\bar C + (1-\bar C)a^{-3(\alpha + 1)} \right)^{-1}.
\label{eq:w_gcg}
\end{equation}

Equation (\ref{eq:w_gcg}) shows that at early times (small $a$, high densities),
$w$ tends
to zero, i.e.\ the gCg behaves as DM,  whereas at later times $w$ tends to $-1$,
i.e.\ the gCg behavior approaches the one expected for DE. It is worth noting
that previous works on the global cosmological dynamics of gCg's dominated
universes have
shown consistency with SNIa, CMB and GRB observations (see e.g.
\cite{bean:2003,colistete:2004,barreiro:2008,liang:2011}).

Past research on linear perturbation theory has shown that although not all
values
of $\alpha$ favor structure formation, there is still some
degree of agreement between gCg UDM and large-scale structure
observations
(see e.g.\ \cite{beca:2003,sandvik:2004,avelino:2004,gorini:2008}). However the
validity of comparing linear theory results with observations
has been questioned recently for the gCg UDM. In particular it has been
noted
in \cite{avelino:2004} that in gCg UDM non-linear effects generate a
backreaction in the background dynamics that cannot be ignored, putting in this
way serious constraints on the validity of linear theory as soon as the first
scales become non-linear.
It is fair to say that non-linear studies are required to state whether or not
the gCg models can become a serious alternative to $\Lambda$CDM.

In \cite{bilic:2004}, the authors have studied the non-linear
evolution 
of dark matter and dark energy in the Chaplygin gas cosmology, using
generalizations 
of the spherical model that incorporated effects of the acoustic
horizon. 
An interesting phenomenon was found there: a fraction of the Chaplygin gas 
condensated
and never reached a stage where its properties changed from dark-matter-like to
dark-energy-like.

A fully non-linear analysis is a cumbersome task usually handled by
hydrodynamical/N-body numerical codes
(see e.g.\ \cite{maccio:2004,aghanim:2009,baldi:2010,li:2011}). 
However, to
best of our knowledge, the gCg case has not yet been addressed,
mainly due to its complex dynamical behavior.

In this paper we focus on the collapse of a spherically symmetric perturbation, with
a classical top-hat profile, leaving the use of N-body techniques to
study the non-linear evolution of gCg perturbations to further
investigations.

We consider a
Friedmann-Lema\^{\i}tre-Robertson-Walker (FLRW) universe with two energetic
components: gCg  and {\it baryons}; since our study is restricted to the
post-recombination epoch we neglect radiation.
Our treatment allows the collapse of both \textit{baryons} and gCg, at variance
with
previous works, see e.g.\ \cite{bilic:2004, multamaki:2004, pace:2010}. We
further assume a time-dependent parameter $w \equiv w(t)$ for both the
collapsing region and the background, and re-examine the definition of the
{\em effective sound speed}  of the perturbed region $c_{eff}^2$. We
derive a more accurate expression for $c_{eff}^2$ rather than using
a approximation as in~\cite{abramo:2008}.

The paper is organized as follows:  Section \ref{sec:SC}
describes the fundamental equations for the SC-TH framework and revises the
notion of {\em effective sound speed}. Section
\ref{sec:numerical} contains our numerical implementation and a discussion of the
results. The paper ends with Section \ref{sec:conclusions} where we draw
our conclusions. An appendix is added for completeness where we
derive the main equations of the SC-TH model.

\section{Spherical `Top Hat' Collapse of Chaplygin gas} \label{sec:SC}

\subsection{The basic equations}

The spherical collapse provides a way to glimpse into the non-linear regime of
perturbation theory, before using more complex methods like N-body simulations.
 Basically, the SC describes the evolution of a spherically symmetric
perturbation embedded in a homogeneous background, which can be static,
expanding or collapsing.
One assumes a spherical `top hat' profile for the perturbed region, i.e.\ a
spherically symmetric perturbation in some region of space with constant
density.
 The assumption of a `top hat' profile
%simplifies even more
further simplifies the SC model as the
uniformity of the perturbation is maintained throughout the collapse,
%, which makes
making its evolution only time dependent.
 As a consequence, we don't need to worry about gradients inside the perturbed
region.
 In essence, the spherical `top-hat' collapse model (SC--TH) describes the
evolution of a homogeneous mini-universe inside a larger
homogeneous universe.

The basic equations used in the SC--TH model are the same that govern
the cosmological background evolution (see e.g.\
\cite{gunn:1972,padmanabhan:1993,mota:2004,abramo:2009}): the
continuity equation
\begin{equation}
\dot\rho = - 3H(\rho+p),
 \label{eq:back_continuity}
\end{equation}
where $H=\dot a/a$ is the Hubble factor, and the Raychaudhuri equation
\begin{equation}
\frac{\ddot a}{a} = -\frac{4\pi G}{3}\sum_i(\rho_i+3p_i)
 \label{eq:back_Raych}.
\end{equation}
where $G$ is the gravitational constant and the $\sum\limits_i(\rho_i+3p_i)$
factor represents the total contribution, of density and pressure, from each
individual component. Please note that in this work we normalize the speed
of light to unity, i.e.\ $c=1$.
For the perturbed region, equations (\ref{eq:back_continuity}) and
(\ref{eq:back_Raych}) are dependent on local quantities and can be written as
\begin{eqnarray}
\dot\rho_c &=& - 3h(\rho_c+p_c)
 \label{eq:coll_continuity}\\
\frac{\ddot r}{r} & = & -\frac{4\pi G}{3}\sum_i(\rho_{c_i}+3p_{c_i})
\label{eq:coll_Raych}.
\end{eqnarray}
where $h=\dot r/r$  and $r $ are respectively the local expansion rate and the
local scale factor;  the perturbed quantities
$\rho_c$ and $p_c$  are related to their background counterparts by:
\begin{eqnarray}
 \rho_c & = & \rho + \delta\rho \\
 p_c & = & p + \delta p.
\end{eqnarray}
A simple relation between $h$ and $H$ can be derived under the SC--TH framework
\cite{abramo:2009},
\begin{equation}
h = H + \frac{\theta}{3a}
 \label{eq:h_rel_H}
\end{equation}
where $\theta \equiv \vec\nabla\cdotp\vec v$ and $\vec v$ is the peculiar
velocity field.

The equations governing the dynamical behavior of
SC--TH are the following \cite{abramo:2009} (see the appendix):
\begin{eqnarray}
\dot\delta_j  & = & -  3H(c^2_{eff_j} -w_j)\delta_j   \nonumber\\
& & -[1+w_j+(1+c^2_{eff_j})\delta_j]\frac{\theta}{a}
\label{eq:dot_delta},\\
\dot\theta & = & - H\theta - \frac{\theta^2}{3a}   \nonumber\\
& & -4\pi Ga\sum\limits_{\rm j}{\rho_{j}\delta_j(1+3c^2_{eff_j})}
\label{eq:dot_theta} ,
\end{eqnarray}
where $\delta_j = (\delta \rho/\rho)_j$ and ${c^2_{eff}}_j
=
(\delta p/\delta \rho)_j$ are, respectively, the density contrast and the square
of the effective sound speed in component $j$
\cite{hu:1998}.

It must be noted that there are as many equations in (\ref{eq:dot_delta})
as the number of cosmological fluid components in the system, while Eq.\
(\ref{eq:dot_theta}) stands alone.
This is true only because we are using a `top-hat' profile, resulting in
$\vec\nabla p = 0$; in a more general situation each cosmological component
would satisfy a corresponding Euler's equation.
It is also convenient to use the scale
factor $a$ instead of the cosmic time $t$ and  the cosmological density
parameters $\Omega_j = \frac{8\pi
G}{3H^2}\rho_j$.
After taking this into account we are left with the following set of equations:
\begin{eqnarray}
\delta_j^{'} & = & -\frac{3}{a}(c^2_{eff_j} - w_j)\delta_j\nonumber\\
& & -[1+w_j+(1+c^2_{eff_j})\delta_j]\frac{\theta}{a^2H}
\label{eq:dot_delta_a},\\
\theta^{'} & = & -\frac{\theta}{a} - \frac{\theta^2}{3a^2H}\nonumber\\
& & -\frac{3H}{2}\sum\limits_{\rm j}{\Omega_j\delta_j(1+3c^2_{eff_j})}
\label{eq:dot_theta_a_omega},
\end{eqnarray}
where the prime denotes the derivative with respect to $a$.

Another useful quantity is the redshift of turnaround $z_{ta}$ which
marks the instant when the perturbed region detaches itself from the background
expansion and starts its collapsing stage (i.e.\ decreasing its physical
radius).
One can define $z_{ta}$ as the redshift at which ${h=0}$, i.e. $z_{ta} =
z({h=0})$.

\begin{table}[t]
\begin{ruledtabular}
\begin{tabular}{llccc}
%\hline
Model & $\alpha$ & $\bar C$  & $z_{ta}$
& $\delta_b(z_{ta})/\delta_{gCg}(z_{ta})$\\
\\ [-3ex]
\hline
a & 0 & 0.75  & 0.138 & 3.02\\
b & $10^{-3}$ & 0.75  & 0.140 & 3.00  \\
c & $10^{-2}$ & 0.75  & 0.168 & 2.83\\
d & $10^{-1}$ & 0.75 & 0.371 & 1.96\\
e & 0.5& 0.75 & 0.685 & 1.20\\
f & 1 & 0.75 & 0.774 & 1.05\\

%\hline
\end{tabular}
\caption{Models used for the evaluation of the SC--TH model in gCg-dominated
universes.}
\label{table:models}
\end{ruledtabular}
\end{table}

\subsection{On the effective sound speed}

We now turn our attention to a decisive player in the dynamics of equations
(\ref{eq:dot_delta_a})--(\ref{eq:dot_theta_a_omega}), the ratio $\delta
p/\delta\rho = c^2_{eff}$.
We argue that a correct computation of $c^2_{eff}$ is crucial to achieve
coherent
results and interpretations.
Under some circumstances $c_{eff}^2$ can be numerically approximated by the
square of the adiabatic sound speed, i.e.\ $c_s^2$,
but they are in fact two different quantities with clear different meanings.
While $c_s^2$ is inherent to the fluid thermodynamics, the factor $c^2_{eff}$
only exists if a perturbation is present.
Moreover, within a perturbed system, it is the value of the ratio $\delta
p/\delta \rho = c_{eff}^2$ (and not $c_s^2 = \partial p/\partial
\rho$), that critically determines the dynamical behavior of the perturbed
region.

The basic equation for the study of the  SC--TH in a gCg-dominated
universe relates the state parameter of the
background $w$ and the parameter relative to the collapsing region  $w_c$ as follows:
\begin{equation}
w_c = \frac{p + \delta p}{\rho + \delta \rho} = \frac{w}{1+\delta} +
c_{eff}^2\frac{\delta}{1+\delta}.
\label{eq:wc_abramo}
\end{equation}
In \cite{abramo:2008}   $w_c$  is
computed by approximating $c_{eff}^2$ by the adiabatic sound speed of the
background (i.e.\ $c_s^2 = \partial p / \partial \rho = -\alpha w$), by trading
 a parameter of the perturbed region for a background one; this is
a good approximation only  if $\delta \ll 1$.
However,  clearly this  is not the case at later stages of the collapse;
in particular  positive values for $w_c$ result out of this approximation,
in disagreement with the equation of state of
the Chaplygin fluid  (see, for instance, Fig.~1 in \cite{abramo:2008}).

Here we remove the
approximation used in \cite{abramo:2008}, and take a step forward
in understanding of the SC in gCg-dominated universes.
This amounts to writing  $c_{eff}^2$ by using the EoS of the
gCg (\ref{eq:EoS_gCg}), and the relation between the densities in the background
and in the collapsed region as follows:
\begin{equation}
 c_{eff}^2 = \frac{\delta p}{ \delta \rho} = \frac{p_c - p}{\rho_c - \rho};
\end{equation}
by using $\rho_c = \rho(1+\delta)$ and Eq.\ (\ref{eq:EoS_gCg}), one obtains

\begin{eqnarray}
 c_{eff}^2 =
& = &-\frac{C}{\rho^{1+\alpha}}\frac{(1+\delta)^{-\alpha}-1}{\delta} =
w\frac{(1+\delta)^{-\alpha}-1}{\delta}\label{eq:my_c2eff} \nonumber \\
\end{eqnarray}

This relation shows an effective sound speed dependent on both the background
and the collapsed region via $w$ and $\delta$ respectively.
It also shows that larger values of $\delta$ will result on smaller values of
$c^2_{eff}$, which is exactly the behavior one should expect in a gCg clump
($\delta \gg 1, w \sim 0, c_s^2 \sim 0$).
Moreover, expanding the term $(1+\delta)^{-\alpha}$ in a Taylor series, for
small values of $\delta$ one can write
\begin{equation}
\label{eq:myCeff_expanded}
c_{eff}^2 = w\frac{1-\alpha\delta+\mathcal{O}(2)-1}{\delta} \simeq -\alpha w,
\end{equation}
thus showing that as $\delta \rightarrow 0$ the value of $c_{eff}^2 \rightarrow
c_s^2$, as expected.
It is always possible to derive an expression for
$c_{eff}^2$, analogous to Eq.\ (\ref{eq:my_c2eff}), for any DE model possessing
an EoS of the type $p=p(\rho)$.

Using the SC--TH framework and Eq. (\ref{eq:EoS_gCg}) it is also possible to
compute directly $w_c$.
Since gCg's state parameter $w$, is given by
\begin{equation}
 w = p/\rho = -\frac{C}{\rho^{1+\alpha}}
\end{equation}
and that $\rho_c = \rho(1+\delta)$, then $w_c$ simply becomes
\begin{equation}
 w_c=-\frac{C}{(\rho(1+\delta))^{1+\alpha}}=\frac{
w}{(1+\delta)^{1+\alpha}}\label{eq:wc_exact}
\end{equation}

Eq.\ (\ref{eq:wc_exact}) clearly shows that for a positive
perturbation, i.e. $\delta > 0$, $w_c$ goes to zero when $\delta$ increases.
The decrease rate of $w_c$ is dependent on the value of $\alpha$, and higher
values of $\alpha$ imply faster variations relative
to the background
value.

\begin{figure}[t]
\includegraphics[width=\columnwidth]{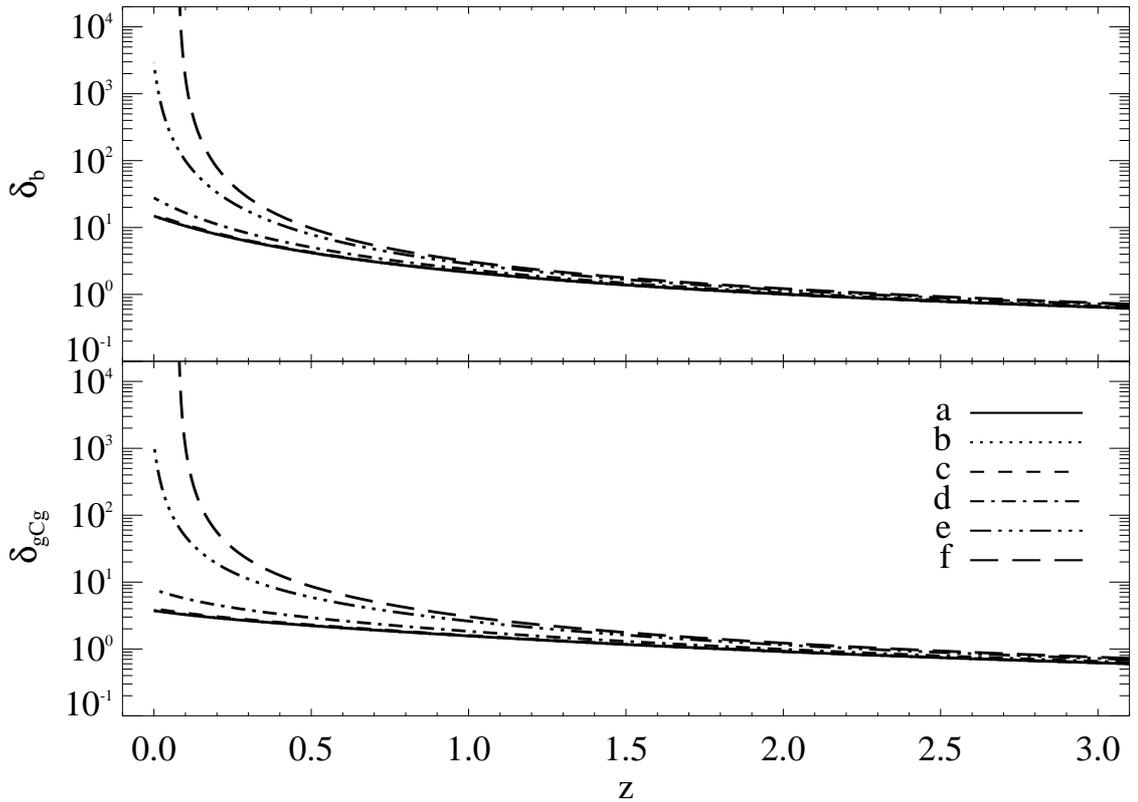}
\caption{Growth of perturbations for the SC -- TH in gCg-dominated
universes.\textit{Top}: $\delta_b$ vs $z$. \textit{Bottom}: $\delta_{gcg}$ vs
$z$}
\label{fig:deltas_vs_z}
\end{figure}

\begin{figure}[t]
\includegraphics[width=\columnwidth]{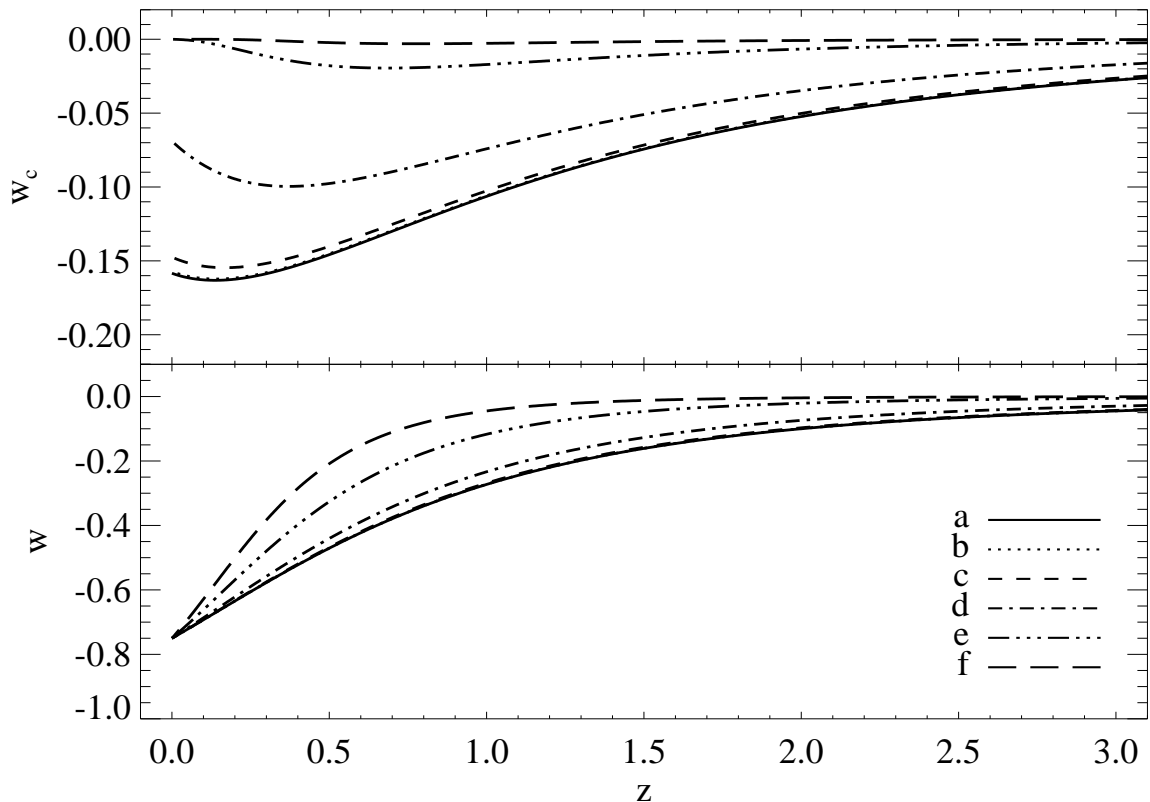}
\caption{gCg's $w_c$ and $w$ evolution with redshift for all the models
presented in Table \ref{table:models}. Higher values of $\alpha$ result
in
values of $w_c$ closer to zero throughout the collapse and in a later
transition
from DM to DE dominated stages of the gCg universes.}
\label{fig:ws_vz_z}
\end{figure}

\begin{figure}[t]
\includegraphics[width=\columnwidth]{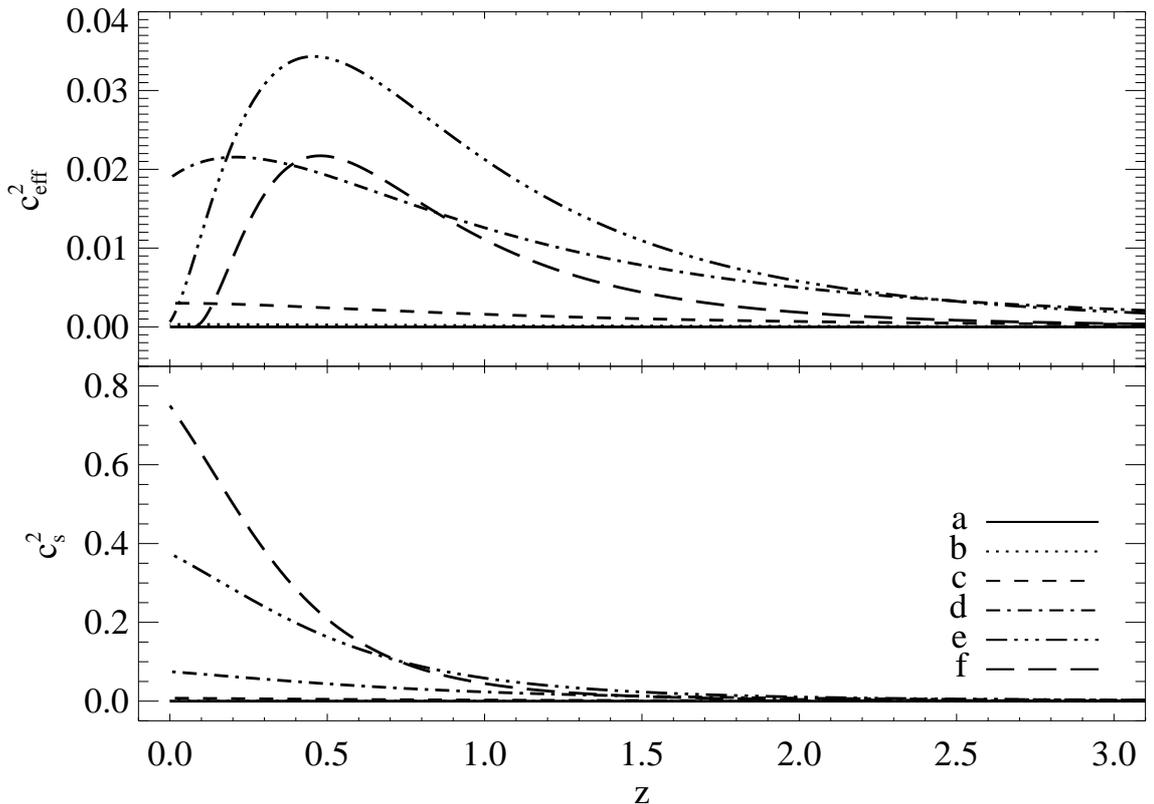}
\caption{gCg's $c_{eff}^2$ and $c_s^2$ evolution with redshift for all
the
models presented in Table \ref{table:models}. Note that, while the value
of
$c_s^2$ increases with the value $\alpha$, the value of $c_{eff}^2$ will
have % a
much more complex dependence.}
\label{fig:ceff2_vz_z}
\end{figure}

\section{The numerical approach} \label{sec:numerical}

\subsection{The method}

 Even though the gCg model has two free parameters, $C$ and $\alpha$, we expect
that $\alpha$ has the largest \textquoteleft
effective\textquoteright\ influence on the growth of
perturbations within the SC--TH framework.
 Not only it is strongly connected to the effective sound speed of
perturbations, but it also dictates for how long the DM stage of the gCg
component lasts.

To investigate the effect of $\alpha$ on the growth of perturbations within
the SC--TH framework, we integrate a system of Ordinary Differential Equations
(ODE), for different values of $\alpha$ while keeping fixed the value of
$\bar C = 0.75$.
The chosen values of $\alpha$ for the numerical integration are shown in Table
\ref{table:models} (recall that model `a' is equivalent to $\Lambda$CDM).
The ODE system is composed by three equations: two of type
(\ref{eq:dot_delta_a}), one for \textit{baryons} and one for gCg, and one of
type (\ref{eq:dot_theta_a_omega}).
The ODE integration was performed by a C++ implementation of a fourth order
Runge-Kutta algorithm from $z=1000$ (at Recombination epoch), to $z=0$ (the
present epoch).
The initial conditions (ICs) for the system are $\delta_{gcg}(z=1000) =
3.5\times10^{-3}$, $\delta_{b}(z=1000) = 10^{-5}$ and $\theta = 0$.
The background is described by a flat FLRW universe with density parameters
$\Omega_{gcg}^0=0.95$ and $\Omega_{b}^0=0.05$, and a Hubble constant $H_0 = 72$
kms$^{-1}$Mpc$^{-1}$, which are consistent with the latest observational values
for the concordance model \cite{komatsu:2011}.

We use a TH profile, so pressure gradients are not present, and we can treat
\textit{baryons} as dust, i.e.
$p_b=w_b=c_{{s}_b}^2=c_{{eff}_b}^2=0$.

\begin{figure}[t]
\includegraphics[width=\columnwidth]{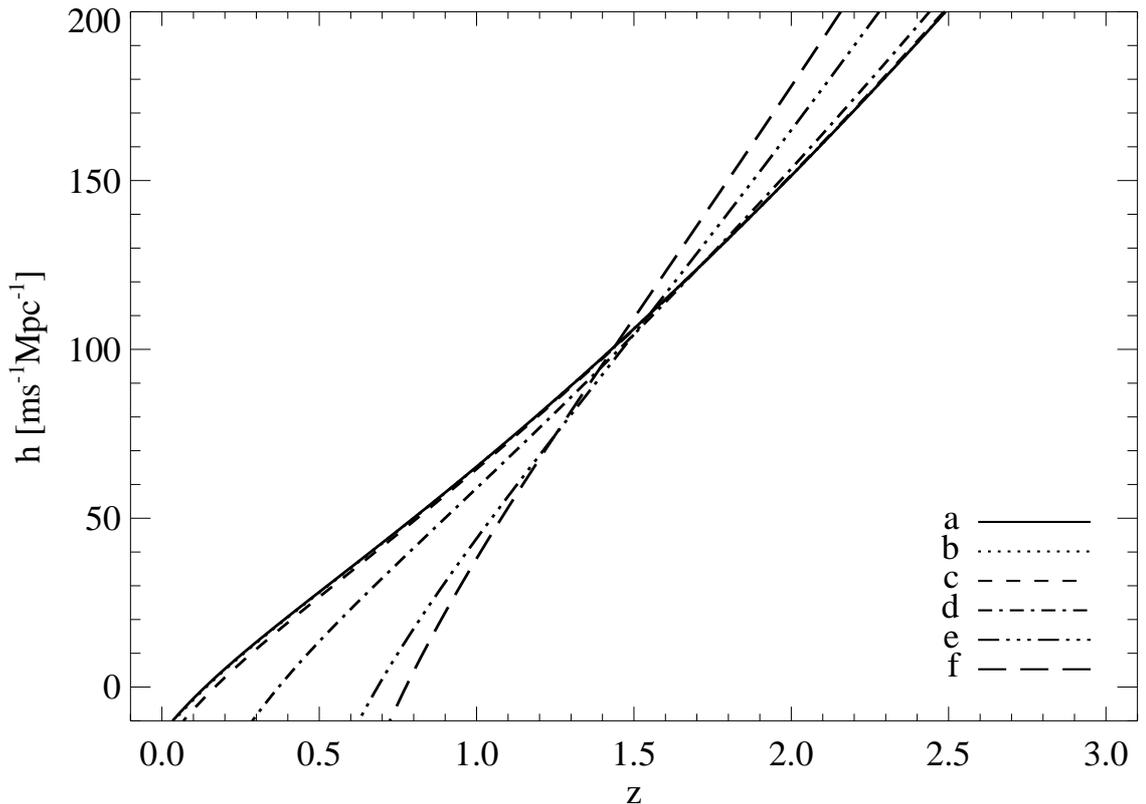}
\caption{Collapsed region expansion rate, $h$, evolution with $z$. Note that
the redshift of
turnaround $z_{ta}$ ($h \simeq 0$) is different for each model and for higher
values of
$\alpha$ it happens at higher redshifts.}
\label{fig:hc_vs_z}
\end{figure}

\subsection{Results and discussion}

When the ICs are the same for all models, the redshift to start the
non-linear regime of the collapse, i.e.\ when $\delta_i \sim 1$, will vary from
model to model.
The different evolutions for $\delta$, shown in Fig.~\ref{fig:deltas_vs_z},
are the result of the $c_{eff}^2$ and $w$ own evolutions and their influence on
equations (\ref{eq:dot_delta_a} --
\ref{eq:dot_theta_a_omega}).
Not only larger values of $\alpha$ will produce faster collapses, via a larger
$c_{eff}^2$ at lower redshifts, but they also extend the DM stage on the gCg
component, thus increasing the collapse in clearly different ways.
As is to be expected, this effect is only visible at later times (when DE
dominates) and, for most of the time, the perturbation evolution remains
undistinguishable between models.

Knowing the $\delta_{gcg}$ evolution with scale factor (or $z$), one can plot
the evolution of $w_c$ using Eq.\ (\ref{eq:wc_exact}) and compare it with the
evolution of $w$.
This comparison, illustrated in Fig.~\ref{fig:ws_vz_z}, is consistent with the
results obtained in Fig.~\ref{fig:deltas_vs_z}, and shows
that indeed gCg changes its dynamical nature throughout the collapse.
Again, one can see that $\alpha$ has a strong imprint in the results, as
higher values of this parameter are connected to a faster collapse, i.e.\ the
value of $w_c$ stays closer to zero throughout the collapse for higher
values of $\alpha$.
Eventually, regardless of the choice of $\alpha$, collapse will occur bringing
$w_c$ close to zero.
Obviously, this is not always true if one increases $\bar C$ too much as that
would be the same as increasing the amount of DE in the system, thus preventing
collapse altogether.

In Fig.~\ref{fig:ceff2_vz_z} we plot the gCg's $c_{eff}^2$ and $c_s^2$ evolution
with redshift.
Clearly, $c_{eff}^2$ has a very different evolution when compared to $c_s^2$
which reinforces the idea that locally the gCg component can behave much
differently than that in the background.
Contrary to what happens with $w_c$ (see Fig.~\ref{fig:ws_vz_z}), the value of
$c_{eff}^2$ has a more complex dependence on $\alpha$.
Another imprint, of the influence of $\alpha$ in the dynamics, can be seen in
the turnaround epochs ($z_{ta}$) for the collapsed region (last column of Table\
\ref{table:models}) where higher values of $\alpha$ stand for an earlier
turnaround.
The turnaround epoch can be defined as the redshift at which $h$ changes sign,
from positive to negative values.
In Fig.~\ref{fig:hc_vs_z}, where we show the evolution of $h$ with redshift,
it is clear that the higher the value of $\alpha$ is, the faster $h$ decreases
and the turnaround point is reached.

It is worth noting that this effect of $\alpha$ in the growth of perturbations
is something akin to the SC--TH model.
Linear perturbation theory in gCg universes has given results that are partially
disagreeing with ours regarding the effect of $\alpha$ in the growth of
perturbations (see e.g. \cite{sandvik:2004,gorini:2008}).
Although this result may seem surprising in fact is something to be expected
as we are using a \textit{top-hat} profile for the density (and by extension
 for the pressure).
As the profile is flat no pressure gradient is present in the dynamics and the
only mechanism that can suppress growth of perturbations is the accelerated
expansion of the universe, which happens only at low redshifts.
This limitation can be alleviated if one uses a continuous like profile for the
initial perturbation, e.g.\ gaussian.
In this case we can fully quantify the influence of $\alpha$ in the formation of
structure, but the use of more complex profiles would imply solving spatial
gradients in the dynamic equations.

% NEW PARAGRAPH FROM REFEREE RECOMENDATION

Finally, we evaluate the influence of ICs in the gCg, ($\delta_{gCg} (z=1000)$),
on the evolution of perturbations, i.e. how the different models in Table
\ref{table:models} compare if one changes the gCg ICs to obtain the same
turn-around redshift for all of them.
We considered the turn-around redshift in Model a, $z_{ta} \simeq$ 0.138, as the
reference one, and force all the others models to turn-around at the same epoch.

The implications on the values of the ICs are summarized in Table
\ref{table:ICsVsZta}.
As can be seen, for the turn-around redshift be the same in all
models, no significant changes in the ICs are required. Even in the worst case
(Model f), ICs differ only by a factor of $\sim$ 0.677, which has no meaning,
in the context of an unrealistic top-hat scenario as this one.

\begin{table}[t]
\begin{ruledtabular}
\begin{tabular}{llccc}
%\hline
Model & $\alpha$ & $\bar C$  & IC ($\times10^{-3}$)
& IC / IC$_a$\\
\\ [-3ex]
\hline
a & 0         & 0.75  & 3.501 & 1\\
b & $10^{-3}$ & 0.75 & 3.498  & 0.999  \\
c & $10^{-2}$ & 0.75 & 3.468  & 0.990 \\
d & $10^{-1}$ & 0.75 & 3.219  & 0.919 \\
e & 0.5       & 0.75 & 2.635  & 0.753\\
f & 1         & 0.75  & 2.370 & 0.677\\
%\hline
\end{tabular}
\caption{Summary for the turn-around normalization, using
the SC-TH framework in gCg-dominated
universes.}
\label{table:ICsVsZta}
\end{ruledtabular}
\end{table}

\begin{figure}[tbh]
\centering
\subfigure[]{\label{fig:ParCurvAlphaX}\includegraphics[
width=0.85\columnwidth]
{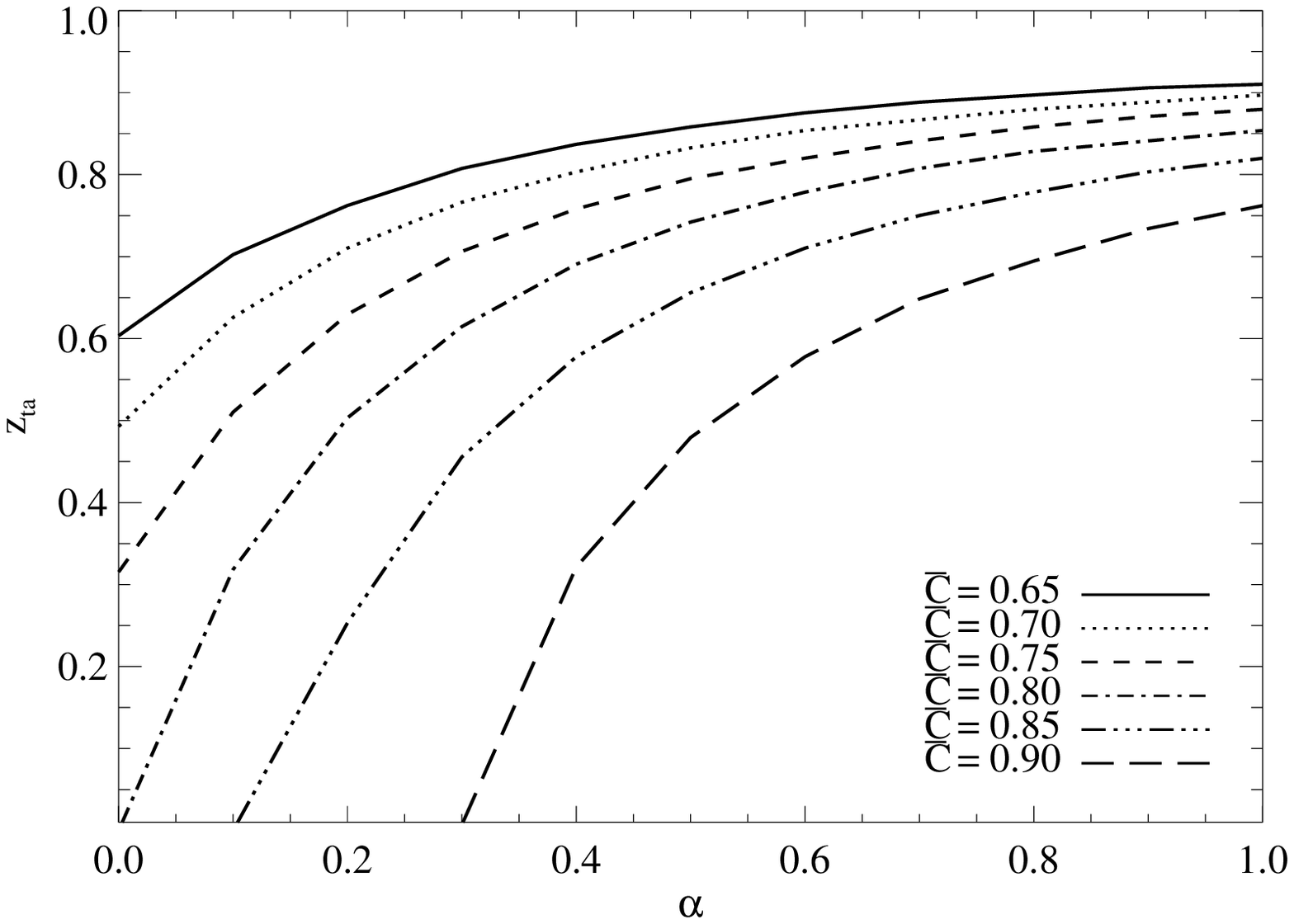}}
\subfigure[]{\label{fig:ParCurvCbarX}\includegraphics[
width=0.85\columnwidth]
{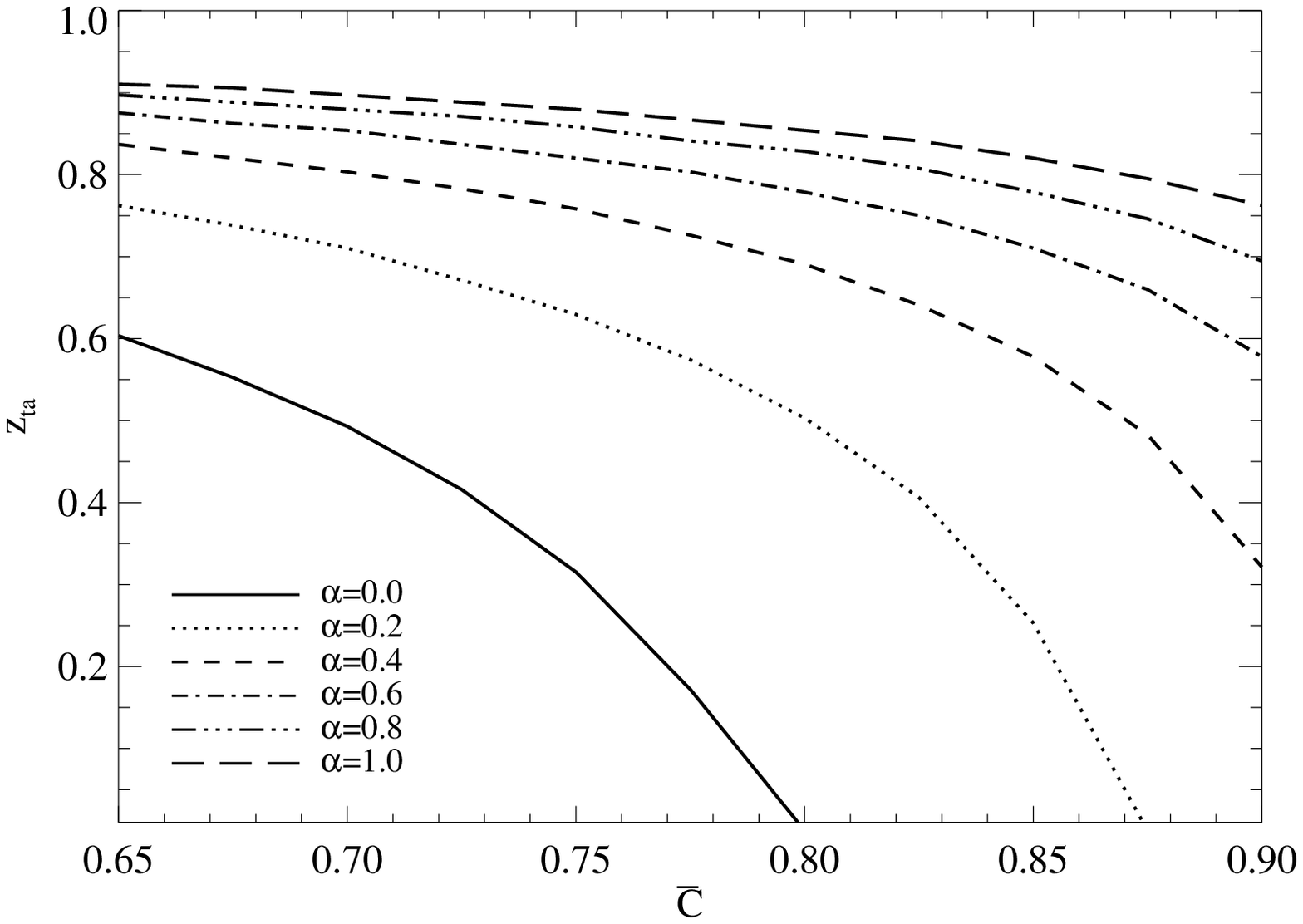}}
\caption{Dependence of $z_{ta}$ on the parameters $\alpha$ and
$\bar C$. \textit{Top:} $z_{ta}$ vs $\alpha$ for different values of $\bar C$.
\textit{Bottom:} $z_{ta}$ vs $\bar C$ for different values of $\alpha$.}
\label{fig:ParCurves}
\end{figure}

\subsection{Combined influence of $\alpha$ and $\bar C$ on the value of
$z_{ta}$}
\label{sec:z_vs_a_C}

To understand how $z_{ta}$ depends on both $\alpha$ and $\bar C$, we constructed
the plots shown in Fig.~\ref{fig:ParCurves}.
As one can see in plot~\ref{fig:ParCurvAlphaX}, $z_{ta}$ is a growing function
of $\alpha$ (independently of the value of $\bar C$), with a more pronounced
growth for higher values of $\bar C$, as expected.
In fact, from Eq.~\ref{eq:w_gcg}, the value of $w_{gCg}$ approaches -1 as $\bar
C$ tends to 1, which implies that the gCg DE-like behavior dominates over its
DM-like behavior, earlier in the system, thus disfavoring the collapse of the
perturbation.

On the other hand, high values of $\alpha$ help gCg to keep its pressure closer
to zero, extending the collapse stage. As a consequence, when $\alpha$ is
large, $z_{ta}$ is almost insensitive to variations of $\bar C$
(see plot~\ref{fig:ParCurvCbarX}).

\section{Conclusions} \label{sec:conclusions}

In this paper we studied the growth of perturbations in
gCg-dominated universes, using the SC--TH model.
The EoS of gCg 
provides an exact relation for the
ratio $\delta p/\delta\rho=c_{eff}^2$ depending on both the redshift and the local density.
Based on this relation we showed that, within the SC--TH framework,
higher values of $\alpha$ speed up the collapse. This result somehow differs
from previous findings obtained in the context of linear theory, e.g.\ in
\cite{beca:2003,sandvik:2004,avelino:2004,gorini:2008}. 
These results are however not directly comparable to ours, which are derived
within the SC--TH framework.

The results obtained here reinforce our expectations on the difference between
the global (linear) and local (non-linear) dynamical behavior for the gCg.
This can be clearly seen in figures \ref{fig:ws_vz_z} and \ref{fig:ceff2_vz_z},
when comparing the evolution of local parameters $w_c$ and
$c_{eff}^2$ to their background analogous $w$ and $c_s^2$.

However, it is worth noting that one should be cautious to take the
conclusions obtained here beyond the SC-TH model limitations.
In particular neglected effects at the perturbation boundary may invalidate the
conclusions \cite{Bilic:2008}.
To better quantify the local
dynamics of gCg-dominated universes as well the impact of local non-linear
inhomogeneities in the background dynamics one would need  
more realistic profiles as well as adequate numerical methods to handle
spatial pressure gradients.
We leave  those developments to future work.

\appendix*
\section{Dynamical Equations}
\label{sec:app_a}
Here we present the derivation of the main equations used in our
study.
First let us consider the difference between Eq. (\ref{eq:coll_continuity})  and
Eq. (\ref{eq:back_continuity}),
\begin{equation}
 \dot\rho_c - \dot\rho=  3H(\rho+p) - 3h(\rho_c+p_c),
\end{equation}
which can be expressed in terms of $w$, $w_c$, $\delta\rho$ and $\delta$,
\begin{equation}
 \frac{d}{dt}(\delta\rho)/\rho=  3H(1+w) - 3h(1+w_c)(1+\delta),
\end{equation}
and by using Eq. (\ref{eq:h_rel_H}) we can also eliminate the variable $h$,
\begin{equation}
  \frac{d}{dt}(\delta\rho)/\rho=  3H(1+w) - 3(H+\theta/3a)(1+w_c)(1+\delta).
\end{equation}
Adding the relation $\delta\dot\rho/\rho=-3H(1+w)\delta$ to both sides of the
equation we get,
\begin{widetext}
\begin{equation}
  \frac{d}{dt}(\delta\rho)/\rho - \delta\dot\rho/\rho=  3H(1+w) -
3(H+\theta/3a)(1+w_c)(1+\delta) + 3H(1+w)\delta,
\end{equation}
that can be written fully in terms of $\dot\delta$ and $\delta$,
\begin{equation}
 \dot{\delta}=  3H(1+w) -
3(H+\theta/3a)(1+w_c)(1+\delta) + 3H(1+w)\delta.
\end{equation}
Recalling that $w_c$ can be expressed in terms of $\delta p$, $w$ and $\delta$
we can recast last equation as,
\begin{equation}
 \dot{\delta}=  3H(1+w) -
3(H+\theta/3a)(1+\frac{w}{1+\delta}+\frac{\delta p}{\rho(1+\delta)})(1+\delta) +
3H(1+w)\delta,
\end{equation}
which then simplifies to,
\begin{equation}
 \dot{\delta} =  3H(1+w)(1+\delta) -
3H(1+\delta+w+\frac{\delta p}{\rho})
-(1+\delta+w+\frac{\delta p}{\rho})\frac{\theta}{a}.
\end{equation}
\end{widetext}
Canceling out the last terms we get,
\begin{equation}
 \dot{\delta} =  3Hw\delta -
3H\frac{\delta p}{\rho}
-\left[1+\delta+w+\frac{\delta p}{\rho}\right]\frac{\theta}{a},
\end{equation}
and finally using $\rho = \delta\rho/\delta$ we arrive to the first dynamical
equation, (\ref{eq:dot_delta}).

To obtain the second dynamical equation we can start from Eq.
(\ref{eq:h_rel_H}), to derive the following equality
\begin{equation}
 \dot h = \dot H + \frac{\dot \theta}{3a} - \frac{\theta H}{3a} =
\frac{\ddot r}{r} - h^2
\end{equation}
 which can be written as,
\begin{equation}
\frac{\ddot r}{r} = \dot H + h^2 + \frac{\dot \theta}{3a} - \frac{\theta H}{3a}.
\end{equation}
Inserting Eq. (\ref{eq:coll_Raych}) into the previous relation we get,
\begin{equation}
\dot H + h^2 + \frac{\dot \theta}{3a} - \frac{\theta H}{3a} = -\frac{4\pi
G}{3}\sum_i(\rho_{c_i}+3p_{c_i})
\end{equation}
where expanding the terms $\dot H$ and $h^2$ we obtain,
\begin{widetext}
\begin{equation}
\frac{\ddot a}{a} - H^2+ H^2+\frac{\theta^2}{3^2a^2} + \frac{2\theta H}{3a}+
\frac{\dot \theta}{3a} - \frac{\theta H}{3a} = -\frac{4\pi
G}{3}\sum_i(\rho_{c_i}+3p_{c_i}).
\end{equation}
\end{widetext}
Finally, removing the extra terms, subtracting the background, and multiplying
by $3a$ on both sides we arrive at
\begin{equation}
\dot\theta +\theta H+ \frac{\theta^2}{3a}=
-4\pi Ga \sum_i(\delta\rho_i+3\delta p_i),
\end{equation}
that can be re-written as Eq. (\ref{eq:dot_theta}) by inserting the relations
$\delta\rho = \rho\delta$ and $c_{eff}^2 =\delta p/\delta\rho$.

\begin{acknowledgments}
We thank F. Hardt, V. Gorini, P.P. Avelino and O. Bertolami for fruitful
discussions. The work of A.K. was partially supported by the RFBR grant
No 11-02-00643. AdS is supported by a Ci\^encia 2007 contract, funded by
FCT/MCTES (Portugal) and POPH/FSE (EC).
\end{acknowledgments}

\bibliography{gcg-exp}

\end{document}